\documentclass[twocolumn,aps,pra,groupedaddress,showpacs,amsmath,amssymb]{revtex4-1}
\usepackage{nicefrac}
\usepackage{siunitx}
\usepackage{mathrsfs}
\usepackage{graphicx} 
\usepackage{hyperref}

\begin{document}

\title{Quasiclassical theory of quantum defect\\ and spectrum of highly excited
rubidium atoms}

\author{Ali Sanayei} 
\affiliation{Institut f\"ur Theoretische Physik}
\affiliation{CQ Center for Collective Quantum Phenomena and their Applications in LISA$^{+}$, Eberhard-Karls-Universit\"at T\"ubingen, Auf der Morgenstelle 14, D-72076 T\"ubingen, Germany}

\author{Nils Schopohl} 
\email[]{nils.schopohl@uni-tuebingen.de}
\affiliation{Institut f\"ur Theoretische Physik}
\affiliation{CQ Center for Collective Quantum Phenomena and their Applications in LISA$^{+}$, Eberhard-Karls-Universit\"at T\"ubingen, Auf der Morgenstelle 14, D-72076 T\"ubingen, Germany}

\author{Jens Grimmel}
\affiliation{Physikalisches Institut} 
\affiliation{CQ Center for Collective Quantum Phenomena and their Applications in LISA$^{+}$, Eberhard-Karls-Universit\"at T\"ubingen, Auf der Morgenstelle 14, D-72076 T\"ubingen, Germany}

\author{Markus Mack} 
\affiliation{Physikalisches Institut} 
\affiliation{CQ Center for Collective Quantum Phenomena and their Applications in LISA$^{+}$, Eberhard-Karls-Universit\"at T\"ubingen, Auf der Morgenstelle 14, D-72076 T\"ubingen, Germany}

\author{Florian Karlewski} 
\affiliation{Physikalisches Institut} 
\affiliation{CQ Center for Collective Quantum Phenomena and their Applications in LISA$^{+}$, Eberhard-Karls-Universit\"at T\"ubingen, Auf der Morgenstelle 14, D-72076 T\"ubingen, Germany}

\author{J\'{o}zsef Fort\'{a}gh}
\email[]{fortagh@uni-tuebingen.de}
\affiliation{Physikalisches Institut} 
\affiliation{CQ Center for Collective Quantum Phenomena and their Applications in LISA$^{+}$, Eberhard-Karls-Universit\"at T\"ubingen, Auf der Morgenstelle 14, D-72076 T\"ubingen, Germany}

\date{\today}

\begin{abstract}
We report on a significant discrepancy between recently published
highly accurate variational calculations and precise measurements
of the spectrum of Rydberg states in $^{87}$Rb on the energy
scale of fine splitting. Introducing a modified effective single-electron
potential we determine the spectrum of the outermost bound electron
from a standard WKB approach. Overall very good agreement with precise
spectroscopic data is obtained. 
\end{abstract}


\maketitle

\section{Introduction}

The spectrum of the outermost bound electron of an alkali atom like
$^{87}$Rb is hydrogen like, but lacks the $n^{2}$-degeneracy
of the eigenstates labeled by the principal quantum number $n$ of
the pure Coulomb potential \cite{Gallagher.1994},\cite{scaled_units}
\begin{equation}
	E_{n,l} = -\frac{1}{\left(n-\delta_{l}\right)^{2}}.
\end{equation}
This effect is the well-known quantum defect $\delta_{l}$, resulting
from the interaction of the outermost electron with the ionic core
of the atom and the nucleus. In a refined version of the statistical
Thomas-Fermi theory \cite{GoeppertMayer.1941}, an effective potential
determining the interaction between the outermost electron and the
nucleus can accurately be modeled by a spherically symmetric potential
$V_{\mathrm{eff}}\left(r;l\right)$ depending on the distance $r$ from
the center and depending on the orbital angular momentum $l\in\left\{ 0,1,2,\ldots,n-1\right\} $
\cite{Greene.1991,Marinescu.1994},\cite{scaled_units}: 
\begin{equation}
	V_{\mathrm{eff}}\left(r;l\right) = -2\left[\frac{Z_{\mathrm{eff}}\left(r;l\right)}{r}+V_{\mathrm{pol}}\left(r;l\right)\right]\label{effective potential Ia}
\end{equation}
Here the function $Z_{\mathrm{eff}}\left(r;l\right)$ represents a
position-dependent weight function that interpolates the value of
the charge between unity for large $r$ and charge number $Z$ near
to the nucleus for $r\rightarrow0$, and $V_{\mathrm{pol}}\left(r;l\right)$
represents a short-ranged interaction taking into account the static
electric polarizability of the ionic core \cite{Born.1925,Gallagher.1994}. 

Overall good agreement with spectroscopic data of alkali atoms (but
discarding the fine splitting) has been reported in \cite{Marinescu.1994}
choosing 
\begin{equation}
	Z_{\mathrm{eff}}\left(r;l\right) = 1+(Z-1)e^{-ra_{1}\left(l\right)}-re^{-ra_{2}\left(l\right)}\left[a_{3}\left(l\right)+ra_{4}\left(l\right)\right]\label{effective potential Ib}
\end{equation}
and
\begin{equation}
	V_{\mathrm{pol}}\left(r;l\right) = \frac{\alpha_{c}}{2}\frac{1-\exp\left[-\left(\frac{r}{r_{c}\left(l\right)}\right)^{6}\right]}{r^{4}}.
\end{equation}
A table of the parameters $a_{1}(l)$, $a_{2}(l)$, $a_{3}(l)$, $a_{4}(l)$,
$\alpha_{c}$, and $r_{c}\left(l\right)$ can be found in \cite{Marinescu.1994}.

In an attempt to also describe the fine splitting of the excitation
spectrum of the outermost electron of $^{87}$Rb, it has
been suggested \cite{Greene.1991} to superimpose 
\emph{a posteriori} a spin-orbit term 
\begin{equation}
	\widetilde{V}_{\mathrm{SO}}\left(r;j,l\right)=\frac{V_{\mathrm{SO}}\left(r;j,l\right)}{\left[1-\alpha^{2}V_{\mathrm{eff}}\left(r;l\right)\right]^{2}},\label{spin-orbit potential Ia}
\end{equation}
on the potential $V_{\mathrm{eff}}\left(r;l\right)$, which then influences the spectrum
$E_{n,j,l}$ on the scale of fine splitting and the orbitals $\psi_{n,j,l}(r)$ accessible to the
outermost electron. Here 
\begin{equation}
	V_{\mathrm{SO}}\left(r;j,l\right)=\alpha^{2}\frac{1}{r}\frac{\partial V_{\mathrm{eff}}\left(r;l\right)}{\partial r}g\left(j,l\right),\label{spin-orbit potential Ib}
\end{equation}
and $\alpha=\frac{\lambda_{C}}{a_{B}}\simeq\frac{1}{137.036}$ denotes
the fine-structure constant, and 
\begin{equation}
	g\left(j,l\right) = 
		\begin{cases}
			0 & \text{ if }l=0,\\
			\\
			\frac{j\left(j+1\right)-l\left(l+1\right)-\frac{3}{4}}{2} & \text{ if }l\geq1,
		\end{cases}
\end{equation}
where $j\in\left\{ l-\frac{1}{2},l+\frac{1}{2}\right\}$. To determine
those orbitals (with principal quantum number $n=n_{r}+l+1$ and radial
quantum number $n_{r}\in\mathbb{N}_{0}$), a normalizable solution
to the Schr\"odinger eigenvalue problem for the radial wavefunction
$U_{n,j,l}(r)=rR_{n,j,l}\left(r\right)$ and associated eigenvalues
$E_{n,j,l}<0$ is required: 
\begin{equation}
	\left[-\frac{d^{2}}{dr^{2}}+\frac{l(l+1)}{r^{2}}+\widetilde{V}\left(r;j,l\right)-E_{n,j,l}\right]U_{n,j,l}(r) = 0,\label{radial Schroedinger eigenvalue problem}
\end{equation}
where 
\begin{equation}
	\widetilde{V}\left(r;j,l\right)=V_{\mathrm{eff}}\left(r;l\right)+\widetilde{V}_{\mathrm{SO}}\left(r;j,l\right)\label{effective potential IIa}
\end{equation}
denotes the effective single-electron potential.

A highly accurate variational calculation of the excitation spectrum
of the outermost electron of $^{87}$Rb has been carried
out recently \cite{Pawlak.2014}, in which the authors expand the
radial wavefunction of the Schr\"odinger eigenvalue problem (\ref{radial Schroedinger eigenvalue problem})
in a basis spanned by $500$ Slater-type orbitals (STOs). On the other
hand, modern high precision spectroscopy 
of Rydberg levels of $^{87}$Rb has been conducted recently. 
Millimeter-wave spectroscopy employing selective field ionization allows for precise measurements of the energy differences between Rydberg levels \cite{Li.2003}. An independent approach is to perform purely optical measurements on absolute Rydberg level energies by observing electromagnetically induced transparency (EIT) \cite{Mohapatra.2007,Mack.2011}. 
However, there is a systematic discrepancy
between variational calculations and the spectroscopic measurements
of the fine splitting 
\begin{equation}
	\Delta E_{n,l}=E_{n,l-\frac{1}{2},l}-E_{n,l+\frac{1}{2},l}\label{fine splitting}
\end{equation}
as shown in Tables \ref{Table I} and \ref{Table II}. Given the
fact that the error bars of the independent experiments \cite{Li.2003,Mack.2011}
are below \SI{1.1}{\mega\hertz} down to \SI{20}{\kilo\hertz}, and on the other hand considering
the high accuracy of the numerical calculations presented in \cite{Pawlak.2014},
such a discrepancy between experiment and theory is indeed significant.

\begin{table*}
	\caption{\label{Table I}Fine splitting $\Delta E_{n,l=1}$ for P states in [\si{\mega\hertz}].}
	\begin{tabular}{|c|c|c|c|c|}
		\hline 
		\textbf{State }$\left\vert n,l=1\right\rangle $  & \textbf{Exp.} \textbf{\cite{Sansonetti.2006}}  & \textbf{\ Exp. \cite{Li.2003}}  & \textbf{Theory \cite{Pawlak.2014}}  & \textbf{Theory (this work)} \\
		\hline 
		8P  & $565.1(4)\times10^{3}$  & N/A  & $602.00\times10^{3}$  & $567.75\times10^{3}$ \\
		\hline 
		10P  & $219.1(4)\times10^{3}$  & N/A  & $231.87\times10^{3}$  & $218.77\times10^{3}$ \\
		\hline 
		30P  & N/A  & $4246.30(5)$  & $4500.50$  & $4246.46$ \\
		\hline 
		35P  & N/A  & $2566.41(32)$  & $2717.41$  & $2566.28$ \\
		\hline 
		45P  & N/A  & $1144.09(13)$  & $1217.24$  & $1143.95$ \\
		\hline 
		55P  & N/A  & $605.77(7)$  & $644.81$  & $605.68$ \\
		\hline 
		60P  & N/A  & $460.76(5)$  & $480.32$  & $460.68$ \\
		\hline 
	\end{tabular}
\end{table*}

\begin{table*}
	\caption{\label{Table II}Fine splitting $\Delta E_{n,l=2}$ for D states in [\si{\mega\hertz}].}
	\centering
	\begin{tabular}{|c|c|c|c|c|c|}
		\hline 
		\textbf{State }$\left\vert n,l=2\right\rangle $  & \textbf{Exp.} \textbf{\cite{Sansonetti.2006}}  & \textbf{Exp.} \textbf{\cite{Li.2003}}  & \textbf{Exp.} \textbf{\cite{Mack.2011}}  & \textbf{Theory \cite{Pawlak.2014}}  & \textbf{Theory (this work)} \\
		\hline 
		8D  & $30.4(4)\times10^{3}$  & N/A  & N/A  & $113.17\times10^{3}$  & $36.42\times10^{3}$ \\
		\hline 
		10D  & $14.9(2)\times10^{3}$  & N/A  & N/A  & $52.05\times10^{3}$  & $16.56\times10^{3}$ \\
		\hline 
		30D  & N/A  & $452.42(18)$  & $452.5(11)$  & $1447.53$  & $456.13$ \\
		\hline 
		35D  & N/A  & $279.65(10)$  & $280.4(11)$  & $894.84$  & $281.52$ \\
		\hline 
		45D  & N/A  & $128.33(4)$  & $127.8(11)$  & $407.64$  & $128.98$ \\
		\hline 
		55D  & N/A  & $69.17(2)$  & $69.4(11)$  & $223.71$  & $69.47$ \\
		\hline 
		57D  & N/A  & $61.98(2)$  & $62.2(11)$  & $197.39$  & $62.24$ \\
		\hline 
	\end{tabular}
\end{table*}

So, what could be the reason for the reported discrepancies? First,
it should be pointed out that in the variational calculations \cite{Pawlak.2014}
a slightly different potential was used, that is, 
\begin{equation}
	V\left(r;j,l\right)=V_{\mathrm{eff}}\left(r;l\right)+V_{\mathrm{SO}}\left(r;j,l\right).\label{effective potential IIb}
\end{equation}
Certainly, within the first-order perturbation theory there exists no noticeable discrepancy 
in the spectrum of the outermost electron on the fine-splitting scale, when 
taking into account the spin-orbit forces
with $V_{\mathrm{SO}}\left(r;j,l\right)$ instead of working with
$\widetilde{V}_{\mathrm{SO}}\left(r;j,l\right)$. This is due to
the differences being negligible for $r>Z\alpha^{2}$. However, since
$V_{\mathrm{SO}}\left(r;j,l\right)$ eventually dominates even the
contribution of the centrifugal barrier term $\frac{l\left(l+1\right)}{r^{2}}$
within the tiny region $0<r\lesssim\alpha^{2}Z$, a subtle problem
with a non-normalizable radial wavefunction $U_{n,j,l}(r)$ emerges when 
attempting to solve the Schr\"odinger eigenvalue problem for any
$l>0$ with the potential $V_{\mathrm{SO}}\left(r;j,l\right)$. Such
a problem is absent when one works with $\widetilde{V}_{\mathrm{SO}}\left(r;j,l\right)$
\cite{Greene.1991}.

A variational calculation with the potential (\ref{effective potential IIb})
employing $N=500$ normalizable STOs as basis functions thus engenders
a systematic (small) error of the matrix elements calculated in \cite{Pawlak.2014}
on the fine-splitting scale. When employing substantially more STOs this error would certainly
become larger. With $N=500$ STOs the discrepancy
of these theoretical results with the high precision spectroscopic
data, as shown in Tables \ref{Table I} and \ref{Table II}, is 
far too large to be corrected by simply replacing
$V_{\mathrm{SO}}\left(r;j,l\right)$ with $\widetilde{V}_{\mathrm{SO}}\left(r;j,l\right)$.
Hence another explanation is required.

\section{Quasiclassical approach and fine splitting of the highly excited $^{87}$Rb}

In 1941 alkali atoms have already been studied in the context of
modern quantum mechanics in the seminal work by Goeppert Mayer \cite{GoeppertMayer.1941},
who emphasized the exceptional role of the $l=1$ and $l=2$ orbitals.
According to Goeppert Mayer, the outermost electron of an alkali atom is governed by
an effective $r$-dependent charge term 
\begin{equation}
	Z_{\mathrm{eff}}\left(r\right)=1+(Z-1)F(r),
\end{equation}
where the function $F(r)$ has been determined by employing the semi-classical
statistical Thomas-Fermi approach to the many-electron-atom problem,
posing the boundary conditions as $\lim_{r\rightarrow0}F(r)=1$ and $\lim_{r\rightarrow\infty}F\left(r\right)=0$.
As discussed by Schwinger \cite{Schwinger.2001}, this approach ceases
to be valid in the inner shell region $Z^{-1}<r<Z^{-\frac{1}{3}}$
of the atom. Therefore, taking into account the fine splitting in
the spectrum of the outermost electron of alkali atoms \emph{a posteriori}
by simply adding the phenomenological spin-orbit term (\ref{spin-orbit potential Ia})
to (\ref{effective potential Ia}), resulting in the effective
single-electron potential (\ref{effective potential IIa}), seems
to be questionable on general grounds in that inner shell region.

On a more fundamental level, the treatment of relativistic effects
in multi-electron-atom spectra requires an \emph{a priori} microscopic
description based on the well-known Breit-Pauli Hamiltonian \cite{Bethe.Salpeter.1957,FroeseFischer.1997}
\begin{equation}
	\mathscr{H}=\mathscr{H}_{\mathrm{nr}}+\mathscr{H}_{\mathrm{rs}}+\mathscr{H}_{\mathrm{fs}}.
\end{equation}
Here $\mathscr{H}_{\mathrm{nr}}$ is the ordinary \emph{nonrelativistic}
many-electron Hamiltonian, while the \emph{relativistic corrections}
are represented by the perturbation operators $\mathscr{H}_{\mathrm{rs}}$
and $\mathscr{H}_{\mathrm{fs}}$. The perturbation term $\mathscr{H}_{\mathrm{rs}}$
contains all the relativistic perturbations like \emph{mass correction},
one- and two-body \emph{Darwin} \emph{terms}, and further the \emph{spin-spin
contact} and \emph{orbit-orbit} terms, which all
commute with the total angular momentum $\mathbf{L}$ and total spin $\mathbf{S}$,
thus effectuating only small \emph{shifts} of the spectrum of the
nonrelativistic Hamiltonian $\mathscr{H}_{\mathrm{nr}}$. The perturbation
operator $\mathscr{H}_{\mathrm{fs}}$ on the other hand breaks the
rotational symmetry. It consists of the standard \emph{nuclear spin-orbit},
the \emph{spin-other-orbit}, and the \emph{spin-spin dipole} interaction
terms, which all commute with $\mathbf{J}=\mathbf{L}+\mathbf{S}$, 
but not with $\mathbf{L}$ or with $\mathbf{S}$ separately, thus
inducing the fine splitting of the nonrelativistic spectrum.

Although the proposed functional form of the potential (\ref{effective potential IIb})
is highly plausible on physical grounds outside the inner core region
$r>Z^{-\frac{1}{3}}$, \emph{prima facie} it appears to be inconsistent
to lump the aforementioned relativistic many-body forces into an effective
single-electron potential of the functional form (\ref{effective potential IIb}),
so that it provides an accurate description also for small distances
$Z^{-1}<r<Z^{-\frac{1}{3}}$.

In the absence of a better microscopic theory for an effective single-electron
potential $V_{\mathrm{eff}}\left(r;j,l\right)$ describing the fine splitting
of the spectrum of the outermost electron in the alkali atoms, we introduce a \emph{cutoff} 
at a distance $r_{\mathrm{so}}(l)$ with $Z^{-1}<r_{\mathrm{so}}(l)<Z^{-\frac{1}{3}}$
so that the effective single-electron potential is
now described by the following modified potential:
\begin{widetext}
\begin{equation}
	\widetilde{V}_{\mathrm{mod}}\left(r;j,l\right) = 
		\begin{cases}
			V_{\mathrm{eff}}\left(r;l\right) & \text{if }0\leq r\leq r_{\mathrm{so}}\left(l\right),\\\\
			V_{\mathrm{eff}}\left(r;l\right)+V_{\mathrm{SO}}\left(r;j,l\right) & \text{if }r>r_{\mathrm{so}}\left(l\right).
		\end{cases}
		\label{effective potential IIc}
\end{equation}
\end{widetext}
The choice \cite{scaled_units}
\begin{align}
	\label{cut off r_so}
	r_{\mathrm{so}}\left(l=1\right) = 0.029483\times r_{c}\left(l=1\right) = 0.0442825,\nonumber \\\\
	r_{\mathrm{so}}\left(l=2\right) = 0.051262\times r_{c}\left(l=2\right) = 0.2495720,\nonumber
\end{align}
gives a surprisingly accurate description of the fine splitting in
the spectroscopic data for all principal quantum numbers $n$, see
Tables \ref{Table I} and \ref{Table II}. 

The calculation of the spectrum of the outermost bound electron is
then reduced to solving the radial Schr\"odinger equation (\ref{radial Schroedinger eigenvalue problem})
with the modified potential $\widetilde{V}_{\mathrm{mod}}\left(r;j,l\right)$.
The resulting spectrum is actually hydrogen like, that is, 
\begin{equation}
	E_{n,j,l} = -\frac{1}{\left(n-\Delta_{j,l}\right)^{2}},\label{quantum defect I}
\end{equation}
where $\Delta_{j,l}$ denotes a quantum defect comprising also the
fine splitting. In actual fact the quantum defect describes a reduction
of the number of nodes $n_{r}$ of the radial wavefunction for $l=0,1,2$
as a result of the short-range interaction of the outermost electron
with the ionic core of the atom. Because the higher the orbital angular
momentum quantum number $l$, the lower the probability of the electron
being located near to the center, it is clear that the quantum defect
decreases rapidly with increasing orbital angular momentum $l$. Therefore,
$\Delta_{j,l}$ is only notably different from zero for $l=0,1,2$. 

Writing $\Delta_{j,l}=\delta_{l}+\eta_{j,l}$ with $\eta_{j,l}\ll\delta_{l}$,
the fine splitting to leading order in $\alpha^{2}$ is:
\begin{equation}
	\Delta E_{n,l} = 2\frac{\eta_{l-\frac{1}{2},l}-\eta_{l+\frac{1}{2},l}}{\left(n-\delta_{l}\right)^{3}}\label{quantum defect I-1}
\end{equation}

The quasiclassical momentum $p\equiv\sqrt{-Q}$ of the bound electron
with orbital angular momentum $l>0$, total angular momentum $j=l\pm\frac{1}{2}$,
and taking into account the Langer shift $l(l+1)\rightarrow\left(l+\frac{1}{2}\right)^{2}$
in the centrifugal barrier \cite{Langer.1937,Berry.1972}, is then given by 
\begin{equation}
	Q\left(r;j,l,E\right) = \frac{\left(l+\frac{1}{2}\right)^{2}}{r^{2}}+\widetilde{V}_{\mathrm{mod}}\left(r;j,l\right)-E.
\end{equation}
For $l=0$ the centrifugal barrier term and the spin-orbit potential
are absent. 

Considering high excitation energies $E<0$ of the bound outermost
electron, i.e. a principal quantum number $n\gg1$, the respective
positions of the turning points $r^{\left(\pm\right)}$ are given
approximately by
\begin{align}
	\label{turning point  l>0}
	r^{\left(-\right)} &= \frac{\left(l+\frac{1}{2}\right)^{2}}{1+\sqrt{1+\left(l+\frac{1}{2}\right)^{2}E}} &  & \text{if }l\geq3,\nonumber \\\\
	r^{\left(+\right)} &\simeq \frac{1}{-E}\left[1+\sqrt{1+\left(l+\frac{1}{2}\right)^{2}E}\right] &  & \text{if }l\geq1,\nonumber
\end{align}
where $0<l\ll\frac{1}{\sqrt{-E}}$. Of course for $l=0$ only a single 
(large) turning point $r^{\left(+\right)}=\frac{2}{-E}$ exists
due to the absence of the centrifugal barrier. However, the lower
turning points $r^{\left(-\right)}$ are strongly modified for $l=1,2$
compared to the pure Coulomb potential case taking into account the
core polarization. For $l=1,2$ the relation 
$r^{\left(-\right)}\left(l\right)\simeq0.02\times r_{c}\left(l\right)$ holds;
that is, $r^{\left(-\right)}\left(l=1\right)\simeq0.03472$ and $r^{\left(-\right)}\left(l=2\right)\simeq0.12827$
\cite{scaled_units}. Since the cutoff $r_{\mathrm{so}}\left(l\right)$
in (\ref{cut off r_so}) is substantially above those values of the
lower turning points $r^{\left(-\right)}\left(l\right)$, a quasiclassical
calculation of the fine-split spectrum of the bound outermost electron
is reliable.

\begin{figure}
	\includegraphics[width=\linewidth]{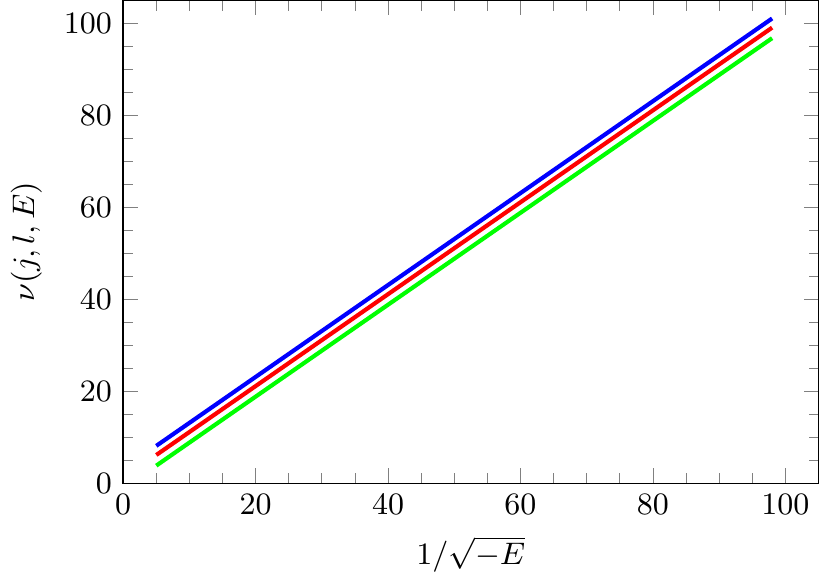}
	\caption{\label{fig:action_integral}(Color online) The action integral $\nu\left(j,l,E\right)$ associated
with the effective single-electron potential $\widetilde{V}_{\mathrm{mod}}\left(r;j,l\right)$
vs. scaled energy $\frac{1}{\sqrt{-E}}$ for $l=0$ (blue line), $l=1$
(red line), $l=2$ (green line), all for $j=l+\frac{1}{2}$. The curves
for $j=l-\frac{1}{2}$ only differ by a tiny shift proportional to
$\alpha^{2}$.}
\end{figure}

For a chosen radial quantum number $n_{r}$, the associated eigenvalues
$E=E_{n,j,l}<0$ of the outermost electron now follow from the WKB
patching condition \cite{Migdal.1977,Karnakov.2013}:
\begin{equation}
	\nu\left(j,l,E\right) \overset{!}{=} 
		\begin{cases}
			n_{r}+1&\text{if }l=0,\\
			\\
			n_{r}+\frac{1}{2}&\text{if }l>0,
		\end{cases}
	\label{WKB patching condition}
\end{equation}
where $\nu\left(j,l,E\right)$ denotes the action integral 
\begin{align}
	\nu\left(j,l,E\right) &= \frac{1}{\pi}\int_{r^{\left(-\right)}}^{r^{\left(+\right)}}\mathrm{d}r\sqrt{-Q\left(r;j,l,E\right)}\nonumber\\
		&= \frac{1}{2\pi}\oint \mathrm{d}r\,p\left(r;j,l,E\right).
\end{align}

Plotting the function $\nu\left(j,l,E\right)$ versus $\frac{1}{\sqrt{-E}}$
for $l=0,1,2$ clearly reveals a linear dependence of the form 
$\nu\left(j,l,E\right)=\frac{1}{\sqrt{-E}}+c\left(j,l\right)$, see Fig. \ref{fig:action_integral}. 

\begin{figure}
	\includegraphics[width=\linewidth]{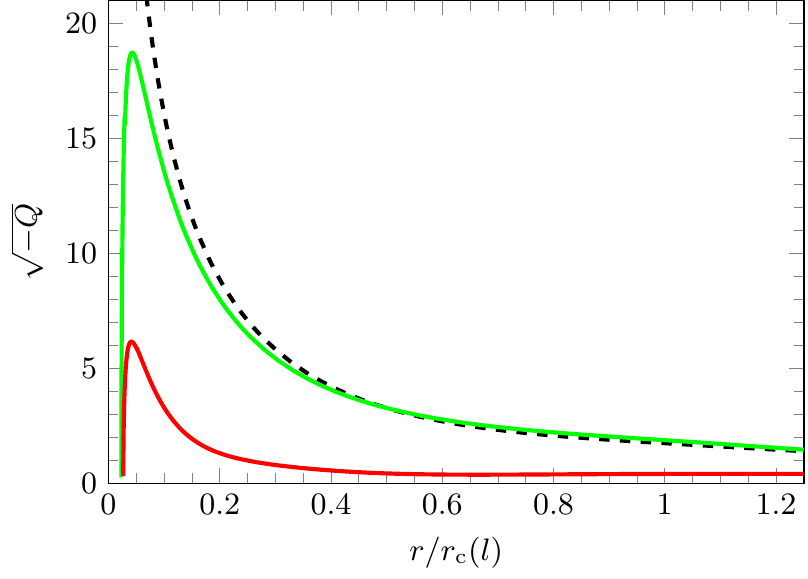}
	\caption{\label{fig:quasiclassical_momentum}(Color online) The quasiclassical momentum $\sqrt{-Q\left(r;j,l,E\right)}$
vs. scaled distance $\frac{r}{r_{c}\left(l\right)}$ for $l=0$ (dashed
black), $l=1$ (green), $l=2$ (red), for $E=E_{n,j,l}$ corresponding
to principal quantum number $n=57$ and $j=l+\frac{1}{2}$. The main
contribution to the quantum defect values in (\ref{quantum defect})
originates from the inner core region $r<r_{c}\left(l\right)$.}
\end{figure}

\begin{table*}
	\caption{\label{Table III}The values of quantum defect $\Delta_{j,l}$ associated with the Rydberg
level $n=57$ for $l=0,1,2$. }
	\begin{tabular}{|c|c|c|c|c|}
		\hline 
		\textbf{Quantum defect $\Delta_{j,l}$} & \textbf{Exp.} \textbf{\cite{Li.2003}}  & \textbf{Exp.} \textbf{\cite{Mack.2011}} & \textbf{Theory \cite{Pawlak.2014}}  & \textbf{Theory (this work)}\\
		\hline 
		$\Delta_{\nicefrac{1}{2},0}$ & $3.1312419(10)$ & $3.13125(2)$ & $3.12791$ & $3.13095$\\
		\hline 
		$\Delta_{\nicefrac{1}{2},1}$ & $2.6549831(10)$ & N/A & $2.65795$ & $2.65197$\\
		\hline 
		\emph{$\Delta_{\nicefrac{3}{2},1}$}  & $2.6417735(10)$ & N/A & $2.64399$ & $2.63876$\\
		\hline 
		\emph{$\Delta_{\nicefrac{1}{2},1}-\Delta_{\nicefrac{3}{2},1}$}  & $0.0132096(14)$ & N/A & $0.01396$ & $0.01321$\\
		\hline 
		\emph{$\Delta_{\nicefrac{3}{2},2}$}  & $1.3478971(4)$ & $1.34789(2)$ & $1.35145$ & $1.34851$\\
		\hline 
		\emph{$\Delta_{\nicefrac{5}{2},2}$}  & $1.3462733(3)$ & $1.34626(2)$ & $1.34628$ & $1.34688$\\
		\hline 
		$\Delta_{\nicefrac{3}{2},2}-\Delta_{\nicefrac{5}{2},2}$ & $0.0016238(5)$ & $0.00163(3)$ & $0.00517$ & $0.00163$\\
		\hline 
	\end{tabular}
\end{table*}

According to \cite{Born.1925}, for $A,B,C,D\in\mathbb{R}$,
with $A>0$, $B>0$, $C>0$, and $\left|D\right|\ll C$ the following
equality holds: 
\begin{equation}
\frac{1}{2\pi}\oint \mathrm{d}r\sqrt{-A+\frac{2B}{r}-\frac{C}{r^{2}}+\frac{D}{r^{3}}} = \frac{B}{\sqrt{A}}-\sqrt{C}+\frac{BD}{2C\sqrt{C}}
\end{equation}
For a pure Coulomb potential $A\equiv-E$, $B\equiv1$, $C\equiv\left(l+\frac{1}{2}\right)^{2}$
and $D\equiv\alpha^{2}g\left(j,l\right)$. The corresponding action
integral then reads
\begin{align}
	\nu^{\left(\mathrm{C}\right)}\left(j,l,E\right) = 
		\begin{cases}
			\frac{1}{\sqrt{-E}} & \text{if }l=0,\\
			\\
			\frac{1}{\sqrt{-E}}-\left(l+\frac{1}{2}\right)+\frac{\alpha^{2}g\left(j,l\right)}{2\left(l+\frac{1}{2}\right)^{3}} & \text{if }l>0.
		\end{cases}
\end{align}
It is thus found from WKB theory that the quantum defect associated
with the single-electron potential $\widetilde{V}_{\mathrm{mod}}\left(r;j,l\right)$
is: 
\begin{equation}
\Delta_{j,l} = \lim_{E\rightarrow0^{-}}\left[\nu\left(j,l,E\right)-\nu^{\left(\mathrm{C}\right)}\left(j,l,E\right)\right]\label{quantum defect}
\end{equation}
Ignoring spin-orbit coupling, i.e. for  $\alpha=0$ , one has $\Delta_{j,l}\equiv\delta_{l}$,
the standard quantum defect. For $l=0$ the centrifugal barrier and
the spin-orbit coupling term (\ref{spin-orbit potential Ib}) are
zero, so $\Delta_{j,l}\rightarrow\Delta_{\frac{1}{2},0}\equiv\delta_{0}$. 

The dependence of the quasiclassical momentum $\sqrt{-Q\left(r;j,l,E\right)}$
on the scaled distance $\frac{r}{r_{c}(l)}$ is shown for
$l=0,1,2$ in Fig. \ref{fig:quasiclassical_momentum}. Clearly, it is the 
inner core region $r^{\left(-\right)}\left(l\right)<r<r_{c}(l)$
that provides the main contribution to the quantum defect values. We find, for $l=0,2$, that changing the fitting
parameter $a_{3}\left(l\right)$ in (\ref{effective potential Ib})
from its tabulated value in \cite{Marinescu.1994} according to the scaling
prescription $a_{3}\left(l=0\right)\rightarrow0.814\times a_{3}\left(l=0\right)$
and $a_{3}\left(l=2\right)\rightarrow0.914\times a_{3}\left(l=2\right)$,
leads to a slight downward \emph{constant} shift of the WKB-quantum
defect. As a result of this change, the calculated WKB-quantum defect
$\Delta_{l\pm\frac{1}{2},l}$ then agrees well with the spectroscopic
data, see Table \ref{Table III}. Such a change of $a_{3}\left(l\right)$
does \emph{not} affect the fine splitting values $\Delta E_{n,l}$
though. We also find that the dependence of the fine splitting $\Delta E_{n,l}$
on the principal quantum number $n$ is well described by (\ref{quantum defect I-1})
for all $n\geq8$, see Tables \ref{Table I} and \ref{Table II}. 

In actual fact, for $r^{\left(+\right)}\gg r^{\left(-\right)}$, which is a
criterion that is always met for high excitation energies $\sqrt{-E}\simeq0$
of the outermost electron, the uniform Langer-WKB wavefunction $U_{n,j,l}^{\left(\mathrm{WKB}\right)}\left(r\right)$
\cite{Langer.1934,Bender.Orszag.1978}, with $r^{\left(+\right)}$ considered as the
only turning point, describes the numerical solution $U_{n,j,l}\left(r\right)$
to the radial differential equation (\ref{radial Schroedinger eigenvalue problem})
under the influence of the effective modified single-electron potential
(\ref{effective potential IIc}) rather accurately \cite{Sanayei.2015}.
Only very near to the second turning point $r^{\left(-\right)}$,
at a distance smaller than $r_{\mathrm{so}}\left(l\right)$, the
Langer-WKB wavefunction $U_{n,j,l}^{\left(\mathrm{WKB}\right)}\left(r\right)$
ceases to be a good approximation to the numerical solution $U_{n,j,l}\left(r\right)$
of the radial Schr\"odinger equation (\ref{radial Schroedinger eigenvalue problem})
\cite{Sanayei.2015}.

\section{Conclusions}

In this work we reported a significant discrepancy between experiment
\cite{Li.2003,Mack.2011} and highly accurate variational calculations
\cite{Pawlak.2014} of the spectrum of Rydberg states of $^{87}$Rb
on the energy scale of the fine splitting. We discussed that the usual
\emph{a posteriori} adding of the relativistic spin-orbit potential
to the effective single electron potential governing the outermost
electron of alkali atoms is indeed inconsistent inside the inner atomic
core region. In the absence of a full microscopic theory that lumps
all many-body interactions together with the
relativistic corrections into an effective single-electron potential in a consistent manner,
we suggested a modified effective single-electron potential, see (\ref{effective potential IIc}),
that enables a correct description of the spectrum of Rydberg states on the fine splitting
scale in terms of a simple WKB-action integral for all principal 
quantum numbers $n\geq8$. Modern precision spectroscopy
of highly excited Rydberg states thus enables the probing of the multi-electron
correlation problem of the ionic core of alkali atoms. This is certainly a fascinating
perspective for further experiments and theoretical studies.

\appendix* 

\begin{acknowledgments}
	This work was financially supported by the FET-Open Xtrack Project HAIRS and the Carl Zeiss Stiftung.
\end{acknowledgments}

\end{document}